\documentclass[10pt,conference,a4paper]{IEEEtran}
\IEEEoverridecommandlockouts
\usepackage{cite}
\usepackage{amsmath,amssymb,amsfonts}
\usepackage{algorithmic}
\usepackage{graphicx}
\usepackage{textcomp}
\usepackage{xcolor}
\usepackage{amsthm}
\usepackage{amssymb}
\usepackage{amsmath}
\usepackage{subfigure}
\usepackage{bm}

\usepackage[linesnumbered, ruled]{algorithm2e}

\def\BibTeX{{\rm B\kern-.05em{\sc i\kern-.025em b}\kern-.08em
    T\kern-.1667em\lower.7ex\hbox{E}\kern-.125emX}}

\begin{document}
\title{RIS-Aided Wireless Amodal Sensing for Single-View 3D Reconstruction\\
}
\author{
Yuhan Wang$^{1}$, Haobo Zhang$^{1}$, Qingyu Liu$^{1}$, Hongliang Zhang$^{2}$, Lingyang Song$^{1,2,3}$\\
$\quad^{1}$School of Electronic and Computer Engineering, Peking University Shenzhen Graduate School, China \\
$\quad^{2}$School of Electronics, Peking University, China\\
$\quad^{3}$Hunan Institute of Advanced Sensing and Information Technology, Xiangtan University, Xiangtan, China\\
\vspace{-0.5cm}}

\maketitle

\begin{abstract}
Amodal sensing is critical for various real-world sensing applications because it can recover the complete shapes of partially occluded objects in complex environments. Among various amodal sensing paradigms, wireless amodal sensing is a potential solution due to its advantages of environmental robustness, privacy preservation, and low cost. However, the sensing data obtained by wireless system is sparse for shape reconstruction because of the low spatial resolution, and this issue is further intensified in complex environments with occlusion. To address this issue, we propose a Reconfigurable Intelligent Surface (RIS)-aided wireless amodal sensing scheme that leverages a large-scale RIS to enhance the spatial resolution and create reflection paths that can bypass the obstacles. A generative learning model is also employed to reconstruct the complete shape based on the sensing data captured from the viewpoint of the RIS. In such a system, it is challenging to optimize the RIS phase shifts because the relationship between RIS phase shifts and amodal sensing accuracy is complex and the closed-form expression is unknown. To tackle this challenge, we develop an error prediction model that learns the mapping from RIS phase shifts to amodal sensing accuracy, and optimizes RIS phase shifts based on this mapping. Experimental results on the benchmark dataset show that our method achieves at least a 56.73\% reduction in reconstruction error compared to conventional schemes under the same number of RIS configurations.

\end{abstract}


\section{Introduction}

3D reconstruction is critical in various real-world applications ranging from autonomous driving to industrial robotics. For example, autonomous vehicles predict the full shape of occluded pedestrians to accurately estimate their position and behavior, and robotic arms need to infer full geometries of stacked goods for efficient grasping~\cite{GridNet}. To tackle the degradation of shape reconstruction accuracy brought by occlusion, amodal sensing techniques have been proposed recently. It infers occluded parts from visible parts by leveraging shape continuity, symmetry, and structural priors of the physical world~\cite{Nanay2018amodal}, thus realizing the reconstruction of complete shapes. 

Among various amodal sensing paradigms, wireless amodal sensing is a potential solution. Specifically, wireless amodal sensing reconstructs complete shapes by analyzing the wireless signals scattered from the target. Compared with existing vision-based amodal sensing methods~\cite{Chu2023DiffComplete}, wireless amodal sensing is robust to lighting and weather variations. It also preserves privacy by not recording sensitive data such as facial details. Compared with LiDAR-based methods~\cite{Yu2023AdaPoinTr}, it is tolerant to adverse weather conditions such as rain and fog, and the hardware cost can be significantly reduced as existing wireless systems such as WiFi can be exploited~\cite{zhu2023echo}.

Despite its potential advantages, wireless amodal sensing faces the issue of data sparsity that limits its performance. Compared to cameras and LiDAR, wireless signals provide lower spatial resolution, leading to sparse sensing data of the target. This issue is further exacerbated in amodal sensing scenarios, where some signal propagation paths are blocked due to occlusion, intensifying data sparsity and making it more difficult to reconstruct the complete shape.

In this paper, we propose a Reconfigurable Intelligent Surface (RIS)-aided wireless amodal sensing scheme to improve the performance of wireless amodal sensing. The proposed system consists of a pair of transmit and receive antennas, an RIS, and a target object to be reconstructed. The deployed RIS consists of massive sub-wavelength elements with tunable phase shifts for beam-steering~\cite{hu2022MetaSketch}. Due to its large aperture, the RIS can enhance the spatial resolution of wireless sensing systems for higher reconstruction accuracy~\cite{tong2021Joint}. In addition, the RIS also creates reflection paths that can bypass the obstacles and partially mitigate occlusions in the environment~\cite{hu2022MetaSketch}. Moreover, a generative model is employed to reconstruct the complete shape based on the sensing data captured from the viewpoint of the RIS.


The proposed RIS-aided amodal sensing scheme is significantly different from existing RIS-aided schemes. Specifically, the authors in~\cite{hu2022MetaSketch} proposed a RIS-aided single-view 3D reconstruction method, where a RIS customizes the reflection beams to capture the information of the target shape. However, this method encounters challenges in reconstructing the backside of objects due to self-occlusion. To alleviate this issue, a multi-view method was designed in~\cite{huang2024Single}, where sensing signals are sent by a moving transmitter that can sense the target from different directions, and the RIS reflects the target-scattered signals to the receiver for shape reconstruction. Nevertheless, this method increases system complexity, and the occlusion may not be fully resolved in complex or dynamic scenarios. Unlike these methods, our scheme leverages a generative model to infer the occluded parts, thereby addressing the occlusion issue and maintaining low system complexity.

The major challenge of RIS-aided wireless amodal sensing is the optimization of RIS phase shift configurations, as the relationship between RIS phase shifts and amodal sensing accuracy is complex, and the closed-form expression is unknown. To address this challenge, we propose a deep neural network (DNN)-based model that approximates the relationship between RIS phase shifts and amodal sensing reconstruction error, enabling effective optimization of RIS phase shift configurations based on this relationship. Experimental results on the benchmark dataset show our method reduces reconstruction error by at least 56.73\% over the baseline with the same number of RIS configurations.



\vspace{-0.3cm}
\begin{figure}[!t]
	\centering
	\includegraphics[width=0.45\textwidth]{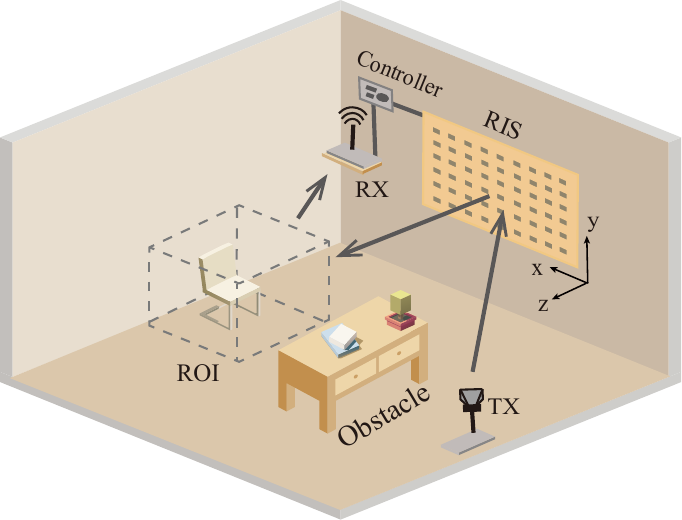}
	\caption{System model of a RIS-aided wireless amodal sensing system.}
	\label{sysmodel}
\end{figure}

\section{System Model}
\subsection{Scenario Description}
As illustrated in Fig.~\ref{sysmodel}, we consider a RIS-aided wireless amodal sensing system comprising a single-antenna transmitter (Tx), a single-antenna receiver (Rx), a target object within the region of interest (ROI), and a vertically deployed RIS with $M$ reconfigurable elements. The presence of other objects between the Tx and the ROI occludes the target object. To fully leverage RIS beamforming, the Tx directs its beam towards the RIS, and only the reflected link via RIS is used for sensing. To enable amodal sensing, the Tx first transmits signals to the RIS, which reflects them toward the object in the ROI. The scattered signals are then received by the Rx and used to reconstruct the object’s complete 3D shape.

\subsection{Object Model}
\label{ROIModel}
\begin{figure}[!t]
	\centering
	\includegraphics[width=0.45\textwidth]{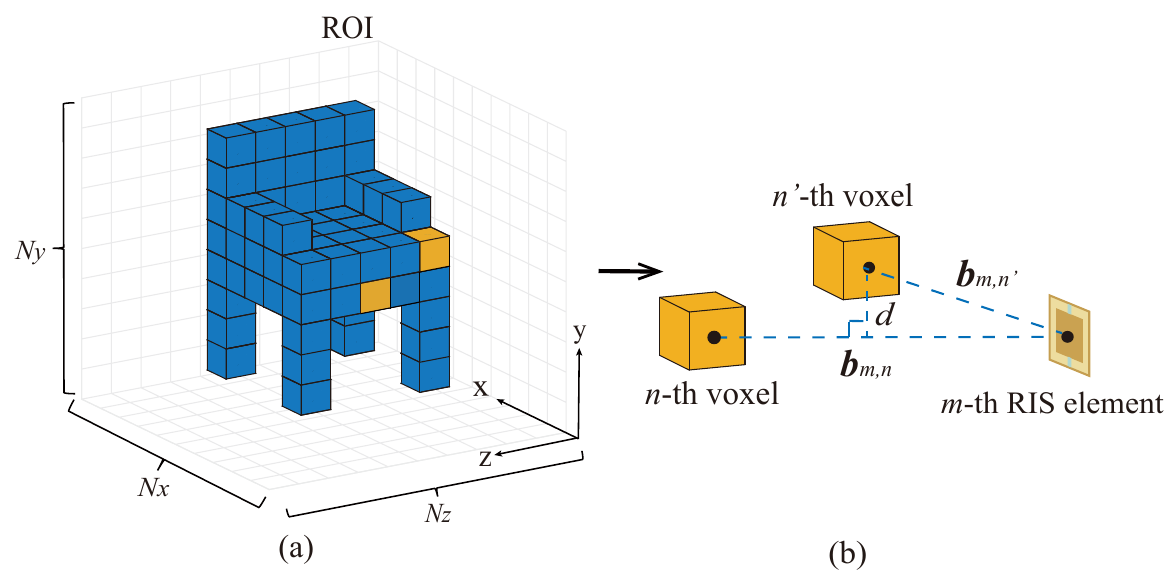}
	\caption{Illustration of the target object and the occlusion among voxels.}

	\label{occlusion}
\end{figure}

As shown in Fig. \ref{occlusion}(a), the ROI is discretized into $N=N_x\times N_y \times N_z$ voxels (each with the size of  $\xi_x\times\xi_y\times\xi_z$). The scattering coefficient of the $n$-th voxel is denoted by $\omega_n$, and the scattering coefficient of the whole ROI can be expressed as
$\boldsymbol{\omega}=[\omega_1, \omega_2, ..., \omega_N]^T$.
Besides, let $\boldsymbol{\chi}=[\chi_1, \chi_2, ..., \chi_N]^T$ denote the shape of the object, where $\chi_n$ represents whether the $n$-th voxel is occupied by the object. Given $\omega_n$, the value of $\chi_n$ can be determined by
\begin{align}
\label{chi}
\chi_n=1\ \text{if}\ \left|\omega_n\right|\geq\tau_{\omega},\text{otherwise}\ 0.
 \end{align} 
where $\tau_{\omega}$ denotes the threshold used to determine whether a voxel is occupied.

From the RIS and Rx perspectives, occupied voxels may block the paths from RIS/Rx to other voxels. Let $v_{m,n}^{RIS}$ and $v_{n}^{Rx}$ denote the blockage (1: unblocked, 0: blocked) for the path from the $m$-th RIS element to the $n$-th voxel, and from the $n$-th voxel to Rx, respectively. These values depend on the locations of other non-empty voxels. Specifically, to block the path from the $m$-th RIS element to the $n$-th voxel, i.e., $v_{m,n}^{RIS}=0$, the location of the non-empty voxel, denoted by the $n'$-th voxel, has to simultaneously satisfy the following conditions~\cite{tong2022Environment}:
\begin{enumerate}
\item{
The distance $d$ between the $n^{\prime}$-th voxel and the path from the $m$-th RIS element to the $n$-th voxel is less than the threshold $\tau_d$, i.e., $d=\left|\bm{b}_{m,n^{\prime}}\times\bm{b}_{m,n}\right|/\left|\bm{b}_{m,n}\right| \textless\tau_d$ as shown in Fig.~\ref{occlusion}(b).}
\item{
The angle between $\bm{b}_{m,n^{\prime}}$ and $\bm{b}_{m,n}$ is acute, i.e., $\bm{b}_{m,n^{\prime}}\cdot\bm{b}_{m,n} \textgreater0$.}
\item{
The projection of $\bm{b}_{m,n^{\prime}}$ onto the direction of $\bm{b}_{m,n}$ is less than the length of $\bm{b}_{m,n}$, i.e., $\left|\bm{b}_{m,n^{\prime}}\cdot\bm{b}_{m,n}\right|\textless\left|\bm{b}_{m,n}\right|^2$.}
\end{enumerate}
The value of $v_{n}^{Rx}$ can be derived similarly.

A voxel is \textit{completely occluded} if all paths from the RIS or Rx to it are blocked, making it invisible to the sensing system. Let the voxel occlusion vector $\bm{v}=[v_1,v_2,...,v_N]^T$ indicate voxel visibility ($v_n=0$: invisible, $v_n=1$: visible). The path occlusion matrix $\bm{V}\in\mathbb{C}^{M\times N}$, with the $(m,n)$-th element being $V_{m,n}=v_{m,n}^{RIS}v_{n}^{Rx}$, determines whether the path from the $m$-th RIS element to the Rx via the $n$-th voxel is blocked. The value of $v_n$ is then given by:
\begin{align}
\label{v_n2}
v_n=0\ \text{if}\ \sum_{m=1}^M V_{m,n}=0,\text{otherwise}\ 1.
 \end{align} 
Then, the object shape $\bm{\chi}$ can be divided into the visible part $\bm{\chi_v}=\bm{\chi}\odot\bm{v}$ and the invisible part $\bm{\chi_i}=\bm{\chi}\odot(\bm{i}-\bm{v})$, where $\bm{i}=[1,1,...,1]^T\in\mathbb{R}^{N\times 1}$ and $\bm{\chi}=\bm{\chi_v}+\bm{\chi_i}$. Similarly, the scattering coefficient of the visible and invisible part are given by $\bm{\omega_v}=\bm{\omega}\odot\bm{v}$ and  $\bm{\omega_i}=\bm{\omega}\odot(\bm{i}-\bm{v})$, respectively.

\subsection{Measurement Model with RIS}
The RIS comprises $M$ elements (size $\xi_s\times\xi_s$), each integrated with PIN diodes that can switch states based on the applied bias voltage. By controlling the states of the PIN diodes, the phase shift applied to the incident signal of each element can be reconfigurable. Assume each element can provide $2^b$ discrete phase shifts, the phase shift of the $m$-th RIS element $\phi_m$ is given by $\phi_m\in\mathcal{F}=\left\{0,\frac{2\pi}{2^b} ,…,\frac{2\pi(2^b-1)}{2^b}\right\}$.

The channel coefficient $h_{ROI}$ of the Tx-RIS-ROI-Rx path can be modeled as the sum of all paths scattered by all RIS elements and ROI voxels~\cite{huang2024Single}:

\small
\begin{equation}
\begin{aligned}
\label{h_kROI1}
&h_{ROI}(\bm{\omega},\bm{q},\bm{V})\\
&=\sum_{m=1}^{M}{\sum_{n=1}^{N}{\frac{\lambda\sqrt{GK_m\xi_s\xi_sG_IF_m^AF_{m,n}^D}}{(4\pi)^{\frac{3}{2}}d_{t,m}^{\alpha}d_{m,n}^{\alpha}d_{n,r}^{\alpha}}e^{-j\frac{2\pi}{\lambda}(d_{t,m}+d_{m,n}+d_{n,r})}}}\\
&\ \ \ \ \times e^{-j\phi_{m}}V_{m,n}\omega_{n},
\end{aligned}
\end{equation}
\normalsize
where $\bm{q}=[e^{-j\phi_{1}},e^{-j\phi_{2}},...,e^{-j\phi_{M}}]^T$ is the RIS phase shift configuration. $\lambda$ is the wavelength. $G$ and $G_I$ correspond to the Tx and RIS element antenna gains, respectively. $K_m$ indicates the Tx antenna's normalized power gain in the direction of the $m$-th RIS element. $F_{m}^A$ and $F_{m,n}^D$ are the normalized power radiation pattern of the $m$-th RIS element for the arrival signal from the Tx and the departure signal to the $n$-th ROI voxel, respectively. $d_{t,m}$, $d_{m,n}$, and $d_{n,r}$ denote the distances between the Tx and the $m$-th RIS element, between the $m$-th RIS element and the $n$-th ROI voxel, and between the $n$-th ROI voxel and the Rx, respectively. $\alpha$ is the path-loss exponent.

Since the Tx-RIS-ROI-Rx path contains the object shape information, its channel gain is first measured from received signals using existing methods~\cite{hu2022MetaSketch, huang2024Single,tong2021Joint} and then employed to reconstruct the object shape. The measurement model of $h_{ROI}$ can be given by $r=h_{ROI}+z$, where $z\sim\mathcal{N}(0,\varepsilon_z)$ is measurement noise~\cite{huang2024Single}. According to (\ref{h_kROI1}), since invisible voxels (with $\sum_{m=1}^MV_{m,n}=0$) do not affect $h_{ROI}$, the measurement reduces to:
\begin{align}
\label{rk1}
r=\bm{q}^T(\bm{H}_r\odot\bm{V})\bm{\omega_v}+z,
\end{align}
where $\bm{H}_r\in\mathbb{C}^{M\times N}$ denotes the free-space channel coefficient of the Tx-RIS-ROI-Rx path.

\subsection{Sensing Protocol}
The proposed RIS-aided amodal sensing protocol operates in three phases:

\textit{1) ROI acquisition: }The ROI's position $(x_{ROI},y_{ROI},z_{ROI})$ and size $(l_x,l_y,l_z)$ are estimated by existing localization techniques~\cite{huang2024Single}. Specifically, the RIS steers reflected beams to scan the entire space, and the Rx analyzes the echo signal strength and time delays from various directions to obtain the ROI position and size.

\textit{2) Measurement: }$K$ RIS configurations are optimized via the method in Section~\ref{RISoptimize}. Then the RIS employs $K$ optimized phase configurations to scan the ROI precisely. For each configuration $k$, the Rx records a measurement $r_k$ of $h_{ROI}$.

\textit{3) Amodal sensing: }The object’s complete 3D shape is reconstructed based on the $K$ measurements using our proposed amodal sensing algorithm detailed in Section \ref{amodal scheme}.

 \section{RIS configuration optimization for Amodel Sensing}
 \label{RISoptimize}
 \subsection{Problem Formulation}
 In this section, we optimize the RIS phase shift configurations to minimize the amodal sensing reconstruction error for both visible and invisible shapes. The optimization problem is formulated as:
   \begin{subequations}\label{qkquestion}
 \begin{align}
 \label{qkquestiona}
 &\min_{\bm{q}_1,\bm{q}_2,...,\bm{q}_K}{a}\\
 \label{qkquestionb}
 ~\text{s.t.}\ 
     &\phi_{k,m}\in\mathcal{F},\forall k\in[1,K],\forall m\in[1,M],
 \end{align}
 \end{subequations}
 where $a$ is the amodal sensing reconstruction error. $\bm{q}_k=[e^{-j\phi_{k,1}},e^{-j\phi_{k,2}},...,e^{-j\phi_{k,M}}]^T$ denotes the $k$-th RIS phase shift configuration with $\phi_{k,m}$ being the phase shift of the $m$-th RIS element. Constraint (\ref{qkquestionb}) requires the phase shift of each RIS element to be selected from the available set $\mathcal{F}$.

 \subsection{Learning-Based RIS Configuration Optimization Algorithm}
  \label{RIS method}
Since the closed-form expression of the amodal sensing reconstruction error is unknown, it is difficult to directly solve the RIS configuration optimization problem. Besides, the channel correlation minimization metric adopted in existing approaches~\cite{zhu2023echo,huang2024Single} cannot be applied for amodal sensing due to the following two reasons: 1) it only focuses on the reconstruction error of the visible shape and ignores the invisible part; 2) voxel occlusion is neglected when calculating the channel correlation. To this end, we train a supervised learning-based error prediction model to learn the mapping from RIS configurations to the amodal sensing reconstruction error, and optimize the RIS configurations accordingly.

\textit{1) Model Architecture:} The proposed error prediction model can estimate the amodal sensing reconstruction error for given RIS configurations. Based on this prediction, the RIS configurations can be optimized through gradient descent to minimize the error. 

Specifically, the model employs a DNN to predict amodal sensing reconstruction error $a$ for given ROI and RIS configurations. It consists of a 7-layer fully connected neural network with layer dimensions being $(L+1)$-32-64-128-64-32-1. The input of the DNN is $L+1$ correlation features extracted from the RIS measurement channel. The RIS configurations $\bm{Q}=[\bm{q}_1,\bm{q}_2,...,\bm{q}_K]^T\in\mathbb{C}^{K\times M}$ and the free-space channel coefficient $\bm{H}_r\in\mathbb{C}^{M\times N}$ are not directly fed into the network because their high dimensions and large configuration space greatly increase the training complexity. Thus, we extract global and $L$ local correlations of the composite channel matrix $\bm{Q}\bm{H}_r$ and train the model with the low-dimensional input to reduce the training complexity, where we have $L<<K,M,N$. The global correlation is defined as~\cite{zhu2023echo}
\begin{align}
\label{c0}
c_0=\Vert R(\bm{Q}\bm{H}_r)-\bm{I}_N\Vert_F/(N^2-N),
\end{align}
where $R(\bm{A})$ denotes the normalized column-wise correlation matrix of matrix $\bm{A}$, with its $(i,j)$-th element given by $R(\bm{A})_{i,j}=|\bm{A}(:,i)^T\bm{A}(:,j)|/\Vert\bm{A}(:,i)\Vert_2\Vert\bm{A}(:,j)\Vert_2$,
representing the correlation between the $i$-th and $j$-th columns of $\bm{A}$. To calculate the local correlation, the ROI is divided into $L$ contiguous cubic sub-regions, each containing $N_l$ adjacent voxels in set $\mathcal{N}_l$. For the $l$-th sub-region, the local correlation is calculated by $c_l=\Vert R(\bm{Q}\bm{H}_r(:,\mathcal{N}_l))-\bm{I}_{N_l}\Vert_F/(N_l^2-N_l)$~\cite{zhu2023echo}. These $L+1$ correlations describe the distinguishability of RIS-to-voxel channel responses across global and local ROI regions~\cite{hu2022MetaSketch}, facilitating the mapping from input to reconstruction error.

\begin{figure}[!t]
	\centering
	\includegraphics[width=0.5\textwidth]{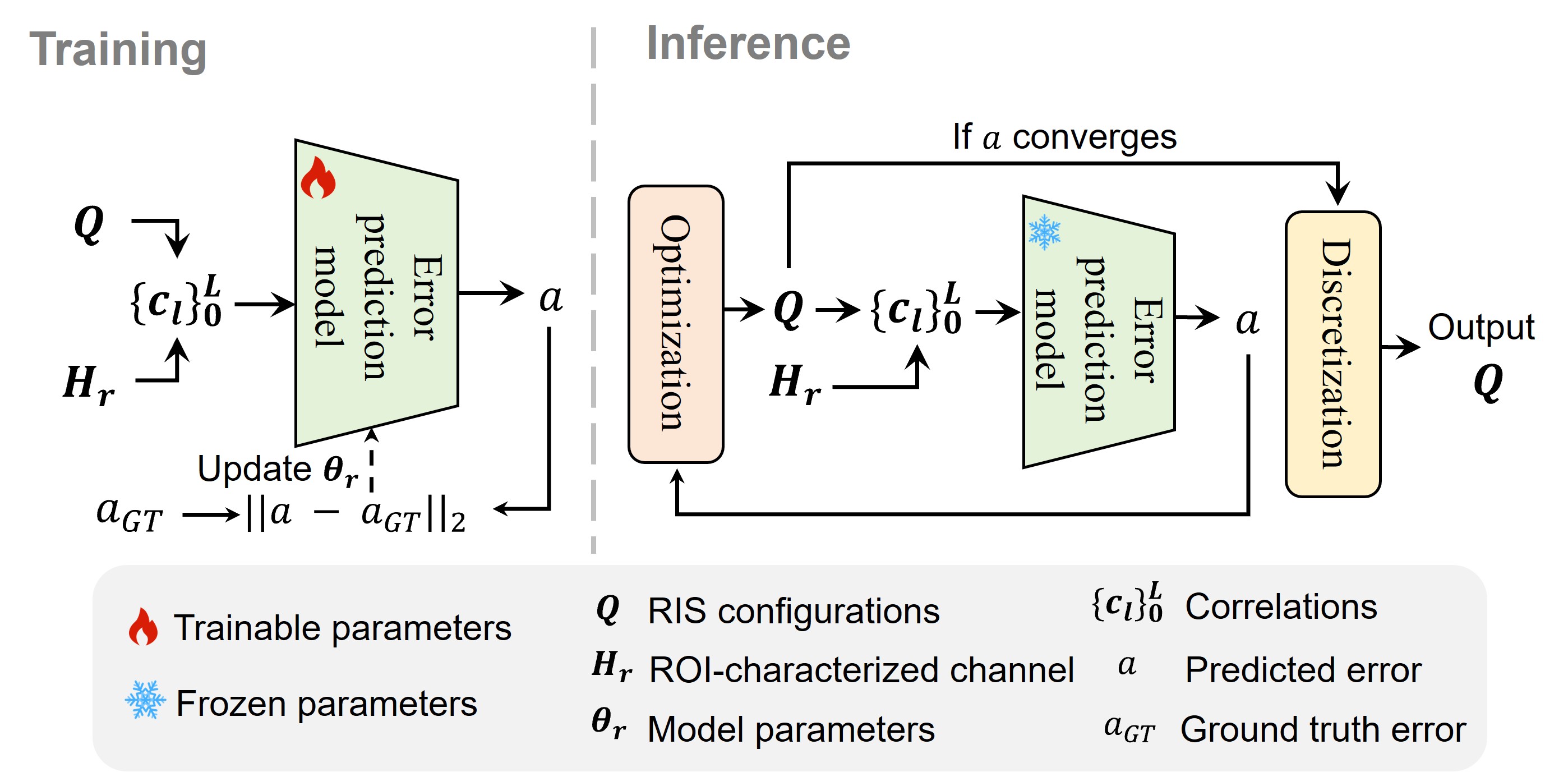}
	\vspace{-0mm}
	\caption{The training and inference of the proposed model for RIS configuration optimization.}
	\label{training}
\end{figure}

\textit{2) Model Training:} We focus on how to train the error prediction model in this part. First, we simulate diverse settings with randomly distributed ROIs and RIS configurations to construct the training dataset. For each setting, the correlations $\{c_l\}_0^L$ are computed. To generate the corresponding ground truth label, we measure and apply the amodal sensing model to a set of ShapeNet~\cite{Chang2015ShapeNet} objects under the same setting, and compute the average reconstruction error across these objects to evaluate the overall reconstruction performance of this setting for arbitrary objects. The error prediction model is finally trained to minimize the mean squared error (MSE) between predicted and true errors, as illustrated in the training phase of Fig.~\ref{training}. 

\textit{3) Model Inference:} As shown in the inference phase of Fig.~\ref{training}, starting from a randomly initialized $\bm{Q}$, $\bm{Q}$ is iteratively optimized based on the predicted error from the pre-trained error prediction model. At each iteration, the current $\bm{Q}$ is first evaluated by the error prediction model to compute the error. Then, the gradient of the predicted error with respect to $\bm{Q}$ is calculated and $\bm{Q}$ is optimized via gradient descent. The iterative process continues until the predicted error converges. After convergence, each element of the optimized $\bm{Q}$ is discretized to its nearest discrete value to satisfy the constraint in (\ref{qkquestionb}).

\section{RIS-aided 3D Reconstruction}
    \label{amodal scheme}
\subsection{Problem Formulation}
In this section, we aim to reconstruct the target object's 3D shape within the ROI. Since measurements $\{r_{k}\}_{k\in[1,K]}$ only capture the visible part according to (\ref{rk1}), we propose to sequentially solve the following two problems. 

\textit{1) Visible shape reconstruction problem: }In this problem, we reconstruct the visible shape from measurements $\{r_{k}\}_{k\in[1,K]}$ to minimize the error between the reconstructed visible shape $\tilde{\bm{\chi_v}}$ and actual visible shape $\bm{\chi_v}$. The problem is given by
\begin{align}
\label{chiquestion1}
\min_{\tilde{\bm{\chi_v}}}{\Vert\tilde{\bm{\chi_v}}-\bm{\chi_v}\Vert_1}, ~\text{s.t.}\ \tilde\chi_{v,n}\in\{0,1\},\tilde{\bm{\chi_v}}=\tilde{\bm{\chi_v}}\odot\bm{v}.
\end{align}
The constraint specifies that the $n$-th element $\tilde\chi_{v,n}$ of $\tilde{\bm{\chi_v}}$ must be binary (0 or 1) and the elements corresponding to invisible voxels must be 0.

 \textit{2) Complete shape reconstruction problem: }In this problem, we recover the complete shape through amodal sensing to minimize the error between the reconstructed and actual shapes, i.e., $\tilde{\boldsymbol{\chi}}$ and $\boldsymbol{\chi}$, respectively. The problem is given by
 \begin{align}
\label{chiquestion2}
\min_{\tilde{\boldsymbol{\chi}}}{\Vert\tilde{\boldsymbol{\chi}}-\boldsymbol{\chi}\Vert_1}, ~\text{s.t.}\ \tilde\chi_{n}\in\{0,1\},\tilde{\bm{\chi_v}}=\tilde{\boldsymbol{\chi}}\ \cap\ \tilde{\bm{\chi_v}}.
\end{align}
The constraint ensures that the $n$-th element $\tilde\chi_{n}$ in $\tilde{\boldsymbol{\chi}}$ is restricted to binary values and the complete shape is reconstructed based on the recovered visible part.

\subsection{Visible Shape Reconstruction Algorithm Design}
\label{Partial}
The visible shape $\bm{\chi_v}$ can be recovered by reconstructing $\bm{\omega_v}$ from $K$ measurements $\{r_{k}\}_{k\in[1,K]}$ through compressed sensing (CS) techniques~\cite{hu2022MetaSketch}. Problem (\ref{chiquestion1}) is first reformulated into a CS problem. Specifically, the CS reconstruction equation can be given by
\begin{align}
\label{rr}
\bm{r}=\bm{Q}(\bm{H}_r\odot\bm{V})\bm{\omega_v}+\bm{z},
\end{align}
where $\bm{r}=[r_1,r_2,...,r_K]^T$. Based on the sparsity of scatters, we can solve $\bm{\omega_v}$ by minimizing the $l_1$-norm of $\bm{\omega_v}$~\cite{hu2022MetaSketch}. Besides, to control the value of the original objective function, we add a constraint in which the error between the reconstructed 
$\bm{Q}(\bm{H}_r\odot\bm{V})\bm{\omega_v}$ and the measurements 
$\bm{r}$ cannot exceed a predefined threshold $\varepsilon_z$, where $\varepsilon_z$ is the variance of $\bm{z}$ in (\ref{rr})~\cite{hu2022MetaSketch}. Therefore, problem (\ref{chiquestion1}) is transformed into the following form~\cite{hu2022MetaSketch}:
\begin{subequations}\label{l1norm}
 \begin{align}
 \label{l1norma}
 &\min_{\bm{\chi_v},\bm{\omega_v}}{\Vert\bm{\omega_v}\Vert_1},\\
 \label{l1normb}
 ~\text{s.t.}\ &\Vert\bm{r}-\bm{Q}(\bm{H}_r\odot\bm{V})\bm{\omega_v}\Vert_2\le\varepsilon_z,\\
  \label{l1normc}
  &\chi_{v,n}=1\ \text{if}\ \left|\omega_{v,n}\right|\geq\tau_{\omega},\text{otherwise}\ 0,\\
\label{l1normd}
&\bm{\omega_{v}}=\bm{\omega_{v}}\odot\bm{v},
 \end{align}
 \end{subequations}
Unlike conventional CS problems, (\ref{l1norm}) involves unknown occlusion variable $\bm{V}$ in the CS measurement matrix $\bm{Q}(\bm{H}_r\odot\bm{V})$, and thus traditional CS methods like GAMP~\cite{Rangan2011Generalized} is not applicable. To this end, we propose an occlusion-aware GAMP algorithm that iteratively solves $\bm{\omega_v}$ and updates $\bm{V}$ based on voxel occlusion relationship mentioned in Section \ref{ROIModel}, enabling accurate recovery of the visible shape $\bm{\chi_v}$.

Let $\bm{\omega_v}^{(i)}$ and $\bm{V}^{(i)}$ denote the solution of $\bm{\omega_v}$ and $\bm{V}$ in the $i$-th iteration, respectively. To initialize, we assume that all the paths are unobstructed, i.e., $\bm{V}^{(0)}$ is all-ones. Then, we iteratively solve $\bm{\omega_v}$ and $\bm{V}$. Specifically, each iteration has the following two steps:

\textbf{Step 1:} Given $\bm{V}^{(i-1)}$, the GAMP algorithm~\cite{Rangan2011Generalized} is executed to derive $\hat{\bm{\omega_v}}^{(i)}$, and the visible scattering coefficient is calculated by $\bm{\omega_v}^{(i)}=\hat{\bm{\omega_v}}^{(i)}\odot\bm{v}^{(i-1)}$ to satisfy constraint (\ref{l1normd}). Here, $\bm{v}^{(i-1)}$ is calculated from $\bm{V}^{(i-1)}$ based on (\ref{v_n2}).

\textbf{Step 2:} Based on $\bm{\omega_v}^{(i)}$, $\bm{V}^{(i)}$ is updated by checking whether the path is blocked by other occupied voxels, as illustrated in Section~\ref{ROIModel}. 

This iterative process continues until $\Vert\bm{V}^{(i)}-\bm{V}^{(i-1)}\Vert_F\le\varepsilon_v$. Finally, the visible shape $\bm{\chi_v}$ is recovered via constraint~(\ref{l1normc}). The visible shape reconstruction is summarized in Algorithm~\ref{visible shape algorithm}.


\begin{algorithm}[!t]
\DontPrintSemicolon
\label{visible shape algorithm}
  \SetAlgoLined
   \KwIn {Measurement $\bm{r}$, configurations $\bm{Q}$, channel $\bm{H}_r$.}
   \KwOut {Visible shape $\bm{\chi_v}$.}

  \textbf{Initialization: } $i=0$, $\bm{V}^{(0)}=\bm{1}_{M\times N}$.\;
  
    \Repeat{$\Vert\bm{V}^{(i)}-\bm{V}^{(i-1)}\Vert_F\le\varepsilon_v$.}{
    $i=i+1$.\;
      Execute GAMP~\cite{Rangan2011Generalized} to obtain $\hat{\bm{\omega_v}}^{(i)}$.\;
    $\bm{\omega_{v}}^{(i)}=\hat{\bm{\omega_{v}}}^{(i)}\odot\bm{v}^{(i-1)}$, where $\bm{v}^{(i-1)}$ is from (\ref{v_n2}).\;
   
    \For{$n=1:N$}{
    Check the path occlusion to determine $v_{m,n}^{RIS}$ and $v_{n}^{Rx}$.\;
    }
        Update $\bm{V}^{(i)}$ based on $V_{m,n}=v_{m,n}^{RIS}v_{n}^{Rx}$.\;
    }
Obtain $\bm{\chi_v}$ according to (\ref{l1normc}).

  \caption{Visible Shape Reconstruction Algorithm}
\end{algorithm}

\subsection{Shape Completion Algorithm Design}
\label{Amodal}
The shape completion problem can be viewed as a conditional generation task, whose goal is to recover the complete 3D shape conditioned on the visible shape. A conditional diffusion model is designed and trained to solve this problem.





\textit{1) Model Architecture:} The conditional diffusion model generates the complete shape $\bm{\chi}_0$ by gradually denoising a random noise shape $\bm{\chi}_T$ through $T$ steps, guided by the condition of visible shape $\bm{\chi_v}$. $\bm{\chi}_t$ denotes the intermediate corrupted shape at the $t$-th step. The model consists of two branches: a main branch that performs the denoising operation, and a control branch that extracts features from the condition input $\bm{\chi_v}$ and injects them into the main branch to guide the denoising process. 

\textbf{Main Branch:} This branch is responsible for iteratively denoising the input corrupted shape. It is built upon the Stable Diffusion framework~\cite{Zhang2020Adding}, adopting a U-Net architecture comprising multiple encoder blocks, a middle block, and multiple decoder blocks, with skip connections linking corresponding encoder and decoder blocks. The input of the main branch is the corrupted shape $\bm{\chi}_{t}$ at the $t$-th step, which is processed to generate the denoised shape $\bm{\chi}_{t-1}$. 

\textbf{Control Branch:} This branch is introduced to process the condition $\bm{\chi_v}$ to guide the denoising process. It shares the same structure as the main branch but uses independent parameters. The input $\bm{\chi_v}$ is processed by its own encoder blocks and middle block to extract multi-scale spatial features. After each block, the extracted features are injected into the decoder block of the main branch through a projection layer via per-voxel additive fusion. This fusion strategy ensures spatial alignment between the features of the visible and generated shapes, and thereby improves shape completion accuracy.

\textit{2) Model Training:} Following the denoising diffusion probabilistic model (DDPM) framework~\cite{Ho2020Denoising}, the model training consists of a forward process and a reverse process. Given a pair of visible and ground truth complete shapes, the complete shape is corrupted into noise through the forward process, and is denoised conditioned on the visible shape via the reverse process. The detail of these two processes is given as follows.

In the forward process, Gaussian noise is gradually added to the complete shape $\bm{\chi}_0$ over $T$ steps to produce a noisy version $\bm{\chi}_T$, with transition $q(\bm{\chi}_t|\bm{\chi}_{t-1})=\mathcal{N}(\sqrt{1-\beta_t}\bm{\chi}_{t-1},\beta_t\bm{I})$, where $\beta_t\in(0,1)$ controls the noise level at each step $t$, and $\bm{\chi}_t$ denotes the corrupted shape result at timestep $t$. 
In the reverse process, a denoising network with learned parameters $\bm{\theta}$ gradually removes the noise from $\bm{\chi}_T$ conditioned on $\bm{\chi_v}$ to recover $\bm{\chi}_0$, with the transition $p_{\bm{\theta}}(\bm{\chi}_{t-1}|\bm{\chi}_t,\bm{\chi_v})=\mathcal{N}(\bm{\mu}_{\bm{\theta}}(\bm{\chi}_t,t,\bm{\chi_v}),\sigma_t\bm{I})$. Here, $\bm{\mu}_{\bm{\theta}}$ represents the network-predicted mean at the current timestep $t$, and $\sigma_t$ denotes the variance. 
Following DDPM~\cite{Ho2020Denoising}, predicting the mean $\bm{\mu}_{\bm{\theta}}$ can be simplified as predicting the noise being removed in the reverse process to approximate the added noise in the forward process. Thus, the model is trained by minimizing the MSE between the predicted noise $\bm{\epsilon}_{\bm{\theta}}(\bm{\chi}_t,t,\bm{\chi_v})$ and truly added noise $\bm{\epsilon}$. 

 \begin{figure*}[!t]
 \subfigure[]{
\begin{minipage}[t]{0.33\linewidth}
\centering
\includegraphics[width=1\textwidth,height=1.8in]{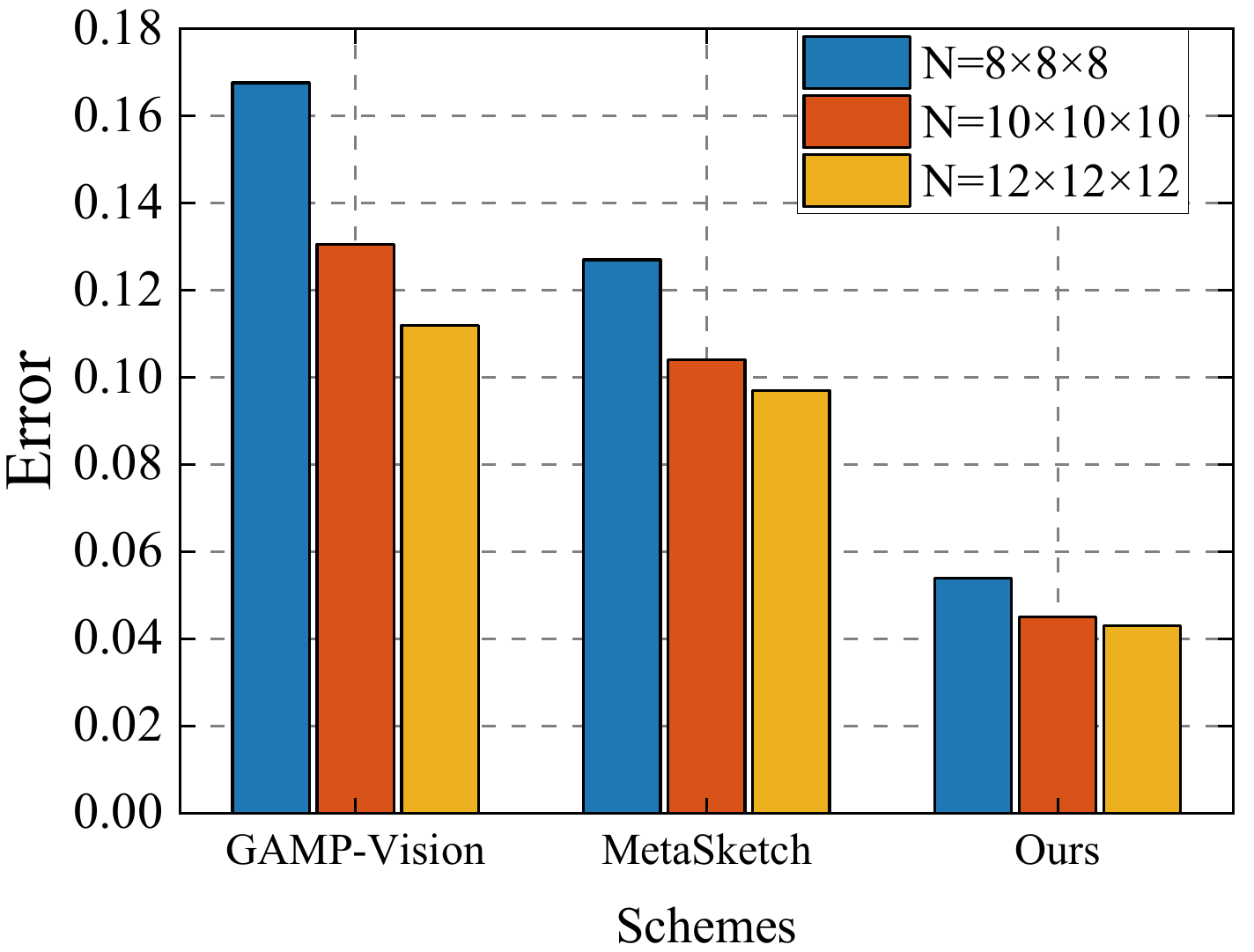}
\label{zongduibi}
\end{minipage}\hspace{-1mm}%
}
\subfigure[]{
\begin{minipage}[t]{0.33\linewidth}
\centering
\includegraphics[width=0.9\textwidth,height=1.8in]{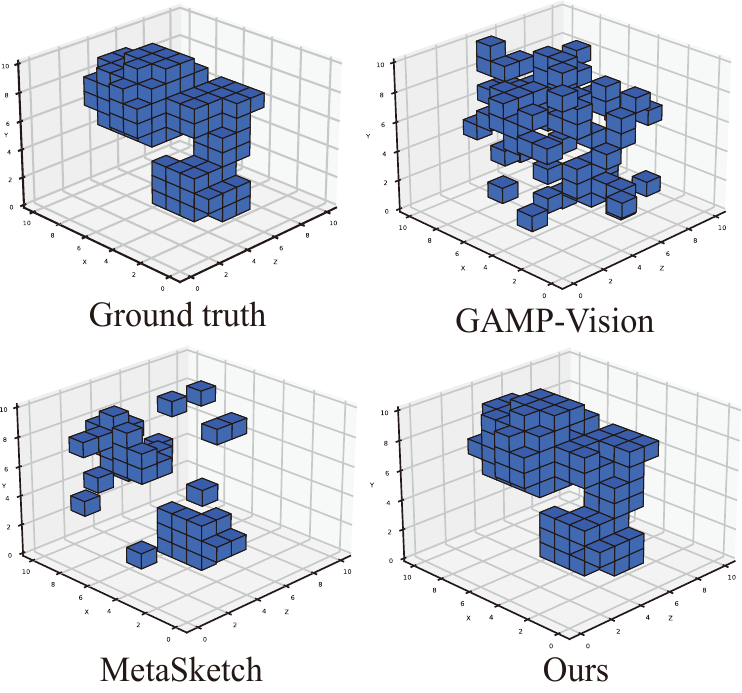}
\label{lamp}
\end{minipage}\hspace{-2mm}
}
\subfigure[]{
\begin{minipage}[t]{0.33\linewidth}
\centering
\includegraphics[width=1\textwidth,height=1.8in]{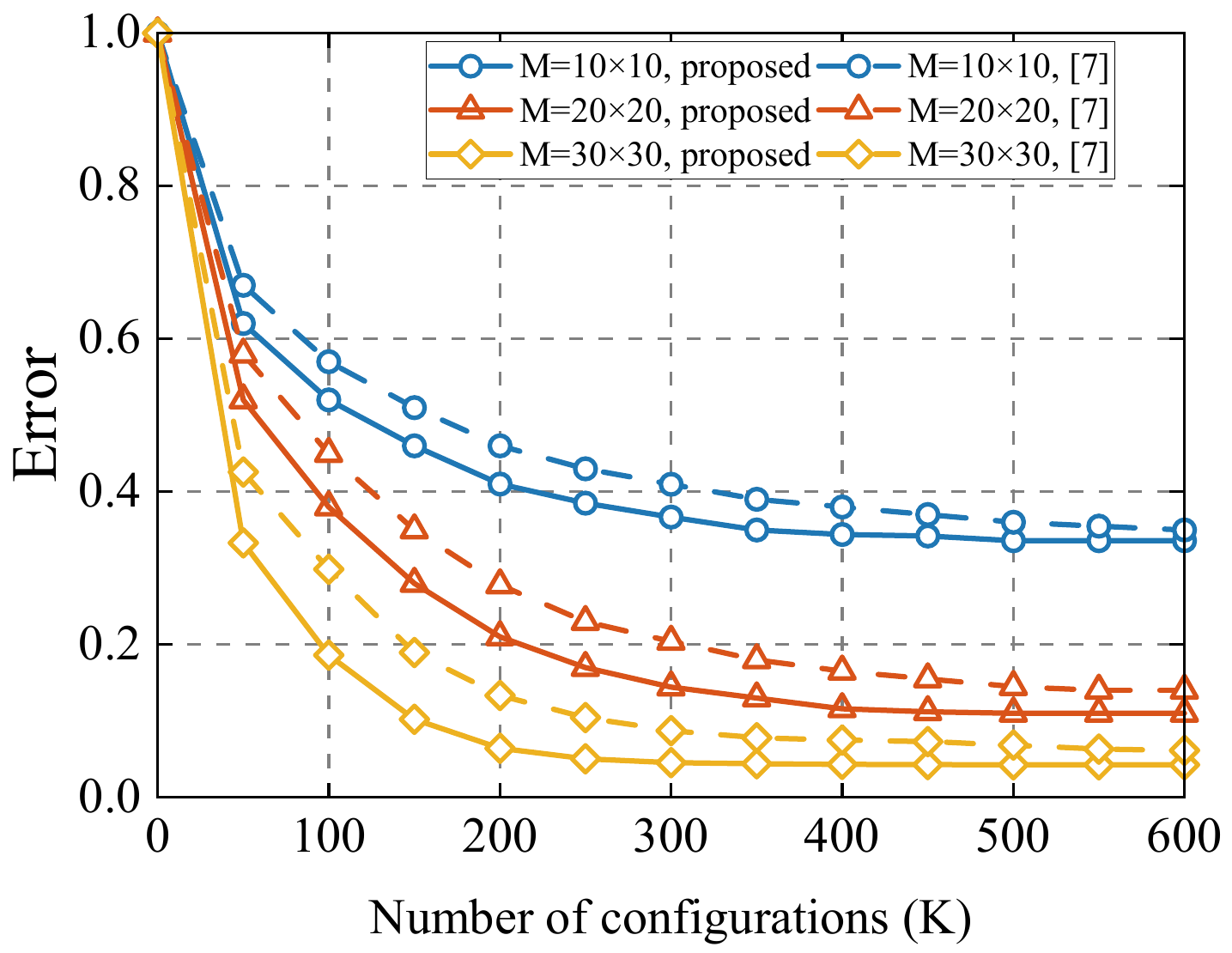}
\label{phase_k}
\end{minipage}
}
\caption{(a) Reconstruction error of the proposed scheme and other sensing schemes. (b) Reconstruction results of a lamp under different schemes (N = 10×10×10). (c) Reconstruction error of our RIS configuration optimization scheme and the correlation minimization-based scheme.}
\vspace{3mm}
\end{figure*}

\section{Experiment Results}
\label{simulation}
In this section, we evaluate the proposed scheme. The RIS lies in the $x$-$y$ plane with its center at the origin, and the Tx and Rx are placed at $(-2.12,0,2.12)\mathrm{m}$ and $(2,0,0)\mathrm{m}$, respectively. The ROI is centered at $(0,0,2.5)\mathrm{m}$ with a size of $3\times3\times3\mathrm{m}^3$, partitioned into $N=10^3$ voxels by default. The Tx transmit power is 35dBm~\cite{hu2022MetaSketch}, and the carrier frequency is $f=3GHz$~\cite{hu2022MetaSketch}. The noise power is $\varepsilon_z^2=-104$dbm~\cite{hu2022MetaSketch}. The distance between RIS elements is 0.05m, and each element has 2-bit phase~\cite{hu2022MetaSketch}. The scattering coefficient is 1 for occupied voxels and 0 for empty voxels. The occlusion thresholds are $\tau_w=0.5$ and $\tau_d=\min\{\xi_x,\xi_y,\xi_z\}$~\cite{tong2022Environment}. Algorithm 1 convergence threshold is $\varepsilon_v=0.01\sqrt{MN}$. For the experiment, 31,000 objects from ShapeNet~\cite{Chang2015ShapeNet} across 10 categories are selected, generating 93,000 targets placed at different viewing angles. To train the shape completion generative model, we generate complete shapes from the targets with a resolution of $32^3$, which supports the completed results to be downsampled to different $N$. The corresponding visible shapes are obtained by simulating RIS observations. The diffusion model uses $T=1000$ steps with $\beta_t$ linearly increasing from $0.0001$ to $0.02$. Based on the trained shape completion generative model, the error prediction model for RIS configuration optimization is then trained as described in Section~\ref{RIS method}.

In Fig.~\ref{zongduibi}, we compare the proposed RIS-aided amodal sensing scheme with two existing baselines: 
\begin{itemize}
    \item GAMP-Vision: an amodal sensing scheme without RIS, and 20\% parts of the objects are occluded. It estimates the visible shape via GAMP~\cite{Rangan2011Generalized} and completes the full shape using the vision-based algorithm~\cite{Chu2023DiffComplete}. 
    \item MetaSketch~\cite{hu2022MetaSketch}: a RIS-aided single-view sensing scheme, which reconstructs scattering coefficients via compressed sensing techniques.
\end{itemize}
It can be observed that the proposed scheme has the lowest errors across different resolutions, and the error is 65.54\% lower than GAMP-Vision and 56.73\% lower than MetaSketch when $N=10\times10\times10$. This is because the wireless link between the object and the transceiver is partially occluded without the aid of RIS in the GAMP-Vision scheme, and MetaSketch cannot recover invisible parts of the object. In contrast, our method exploits RIS to bypass the occlusion and employs a generative model to infer the invisible part, thus having a lower 3D shape reconstruction error.

Fig.~\ref{lamp} presents a visual example of object shape reconstruction obtained by different schemes. Due to occlusion and the lack of RIS assistance, the GAMP-Vision result shows severe errors with the object shape being barely recognizable. While the RIS-aided MetaSketch can reconstruct the visible part, it fails to recover the invisible part, leading to incomplete and inaccurate 3D reconstruction. In contrast, our proposed method produces a reconstruction result that closely matches the ground truth, effectively capturing both visible and invisible parts of the object.

Fig.\ref{phase_k} shows the reconstruction error versus the number of RIS configurations to evaluate the performance of the proposed RIS configuration optimization method. For comparison, we also provide the results obtained by minimizing the channel correlation \cite{zhu2023echo} to optimize RIS configurations rather than using the proposed supervised learning-based method. As the number of configurations $K$ increases, the reconstruction error decreases and gradually converges. The error also decreases when the RIS size $M$ grows, and the performance gain becomes smaller as $M$ continues to increase. Across all settings, our method consistently outperforms the correlation minimization-based scheme. Notably, for fixed $M$, our method achieves a lower error with fewer configurations compared to the existing scheme, highlighting its efficiency by considering the amodal sensing reconstruction error of both visible and invisible shapes.

 \section{Conclusion}
In this paper, we have considered an RIS-aided amodal sensing system, which recovers the complete shape of partially occluded objects with the aid of a large-scale RIS and a generative AI model. Given that the relationship between RIS configurations and amodal sensing reconstruction error is complex and lacks a closed-form expression, we proposed a learning-based RIS configuration optimization scheme that approximates this relationship and optimizes RIS phase shift configurations accordingly. Experiment results lead to two key conclusions:
\begin{enumerate}
\item{Compared to existing single-view sensing schemes, the proposed RIS-aided amodal sensing scheme reduces reconstruction error by at least 56.73\%.
}
\item{The reconstruction error decreases as the resolution of the ROI and the number of RIS configurations increase.}
\end{enumerate}

\section*{Acknowledgement}

This work was supported in part by the National Science Foundation under Grant 62371011; in part by the Beijing Natural Science Foundation under Grant L243002; in part by the GuangDong Basic and Applied Basic Research Foundation under Grant 2023B0303000019; in part by the Shenzhen Science and Technology Program under Grant JCYJ20241202125910015; and in part by the Beijing Outstanding Young Scientist Program JWZ020240102001.




\vspace{0.5cm}
\begin{thebibliography}{00}
\bibitem{GridNet}
Y. Yue \emph{et al.}, ``GridNet-3D: A novel real-time 3D object detection algorithm based on point cloud,'' \emph{Chin. J. Electron.}, vol.~30, no.~5, pp. 931--939, Sep. 2021.

\bibitem{Nanay2018amodal}
B. Nanay, ``The importance of amodal completion in everyday perception,'' \emph{i-Perception}, vol. 9, no. 4, p. 2041669518788887, Jul. 2018.
\bibitem{Chu2023DiffComplete}
R. Chu, E. Xie, S. Mo \emph{et al.}, ``DiffComplete: Diffusion-based generative 3D shape completion,'' in \emph{Adv. Neural Inf. Process. Syst. (NeurIPS)}, \hskip 1em plus 0.5em minus 0.4em\relax LA, USA, Dec. 2023, pp. 75951--75966.
\bibitem{Yu2023AdaPoinTr}
X. Yu \emph{et al.}, ``AdaPoinTr: Diverse point cloud completion with adaptive geometry-aware transformers,'' \emph{IEEE Trans. Pattern Anal. Mach. Intell.}, vol. 45, no. 12, pp. 14 114--14 130, Dec. 2023.

\bibitem{hu2022MetaSketch}
J. Hu, H. Zhang, K. Bian \emph{et al.}, ``MetaSketch: Wireless semantic segmentation by reconfigurable intelligent surfaces,'' \emph{IEEE Trans. Wireless Commun.}, vol. 21, no. 8, pp. 5916--5929, Aug. 2022.

\bibitem{tong2021Joint}
X. Tong \emph{et al.}, ``Joint multi-user communication and sensing exploiting both signal and environment sparsity,'' \emph{IEEE J. Sel. Top. Signal Process.}, vol. 15, no. 6, pp. 1409--1422, Nov. 2021.


\bibitem{zhu2023echo}
S. Zhu, Z. Yu, Q. Guo, J. Ding, Q. Cheng, and T. J. Cui, ``RIS assisted joint uplink communication and imaging: Phase optimization and bayesian echo decoupling,'' 2023, arxiv:2301.03817.
\bibitem{huang2024Single}
Y. Huang, J. Yang, C. K. Wen and S. Jin, ``RIS-sided single-frequency 3D imaging by exploiting multi-view image correlations,'' \emph{IEEE Trans. Commun.}, vol. 72, no. 8, pp. 5003--5018, Aug. 2024.


\bibitem{tong2022Environment}
X. Tong \emph{et al.}, ``Environment Sensing Considering the Occlusion Effect: A Multi-View Approach,'' \emph{IEEE Trans. Signal Process.}, vol. 70, pp. 3598--3615, Jun. 2022. 


\bibitem{Chang2015ShapeNet}
A. X. Chang \emph{et al.}, ``ShapeNet: An information-rich 3D model repository,'' \emph{arXiv preprint arXiv:1512.03012}, 2015.
\bibitem{Rangan2011Generalized}
S. Rangan, ``Generalized approximate message passing for estimation with random linear mixing,'' \emph{Proc. IEEE Int. Symp. Inf. Theory Proc.}, 2011, pp. 2168--2172.
\bibitem{Ho2020Denoising}
J. Ho, A. Jain, P. Abbeel, ``Denoising diffusion probabilistic models,'' \emph{Adv. Neural Inf. Process. Syst.}, 2020, pp. 6840--6851.
\bibitem{Zhang2020Adding}
L. Zhang, A. Rao, M. Agrawala, ``Adding conditional control to text-to-image diffusion models,'' \emph{Proc. IEEE/CVF Int. Conf. Comput. Vis.}, 2023, pp. 3836--3847.




\end{thebibliography}
\end{document}